\documentstyle[twocolumn,aps,pra,psfig,epsf,epsfig]{revtex}
\draft
\def\be{\begin{equation}}
\def\e#1{\label{#1}\end{equation}}
\def\bea{\begin{eqnarray}}
\def\ea#1{\label{#1}\end{eqnarray}}

\def\ee{\end{equation}}
\def\eea{\end{eqnarray}}
\def\bem#1{\begin{mathletters}\label{#1}}
\def\eml{\end{mathletters}}
\def\br{{\bf r}}
\def\bk{{\bf k}}

\begin{document}
\title{``Supersolid'' self-bound Bose condensates via laser-induced
interatomic forces}
\author{S. Giovanazzi \and D. O'Dell and G. Kurizki \\
 Weizmann Institute of Science,
76100 Rehovot, Israel }
\date{\today}
\maketitle

\begin{abstract}
We show that the dipole-dipole interatomic forces induced by a single 
off-resonant running laser beam can lead to a self-bound pencil-shaped
Bose condensate, even if the laser beam is a plane-wave.
For an appropriate laser intensity the ground state has a 
quasi-one dimensional density modulation --- a Bose ``supersolid''.
\pacs{PACS: 03.75.Fi, 34.20.Cf, 34.80.Qb, 04.40.-b}
\end{abstract}

Electrostriction is the tendency of matter to become
compressed in the presence of an electric field \cite{pita}. 
In an optically trapped atomic Bose-Einstein condensate (BEC) 
it is provided by the gradient of the incident electric field 
(one-body dipole forces). 
However, it can occur even if the external fields are \emph{homogeneous}:
the electrostriction effects of dipole-dipole
interatomic forces induced by certain configurations of 
far-off-resonant laser beams are capable, 
for sufficiently high laser intensity, of self-binding a BEC within 
a wavelength of the laser (in the near-zone) \cite{giova2001a}.
Thereby, a physical situation can be realized which has analogies to 
self-gravity \cite{odell2000,giova2001b}.

Here we show that even in the simplest case of a \emph{single} 
homogeneous, circularly-polarized, far-off-resonant 
laser beam a BEC can be \emph{self-bound} by electrostriction,
due to the retarded part of the dipole-dipole interaction.
The self-bound condensate will be pencil shaped, 
aligned along the direction of propagation of light,
in principle even for arbitrarily small intensity.
As the laser intensity increases, the electrostriction
can give rise to a remarkable density modulation of the quasi-one dimensional
condensate ground state. 
Such a condensate bears similarity to a ``supersolid'', i.e., 
a long-range crystalline-like density modulation imposed
upon a superfluid by interparticle forces \cite{pomeau94}.
The formation of such structures is associated with
a strong enhancement of the elastically scattered field
(and, consequently, a large radiation pressure) 
and with  suppression of the heating due to spontaneous Rayleigh scattering.
The relation of our effects to (stimulated) Brillouin scattering,
collective atomic recoil and superradiant Rayleigh scattering 
\cite{inouye,moore1999} is briefly discussed.

We consider the case of a far-off-resonant circularly polarized 
laser beam propagating along the positive $\hat{z}$ direction
with wavevector $q$ (Fig.\ 1---inset).
The dipole-dipole induced interatomic potential 
\cite{thirunamachandran80} then becomes 
\begin{eqnarray}
V_{\mathrm{dd}}(\vec{r}) &=&
\frac{ 3 }{4 }\hbar \Gamma_{ray}
\Big[ \frac{2 z^2 - x^2 - y^2 }{ q^3 r^5} 
\big( \cos q r +qr \sin q r \big) 
\nonumber \\
&-& \frac{ 2 z^2 + x^2 + y^2 }{ q r^3} 
\cos qr \Big] \cos (q z)   \label{eq:ztpot}
\;.
\label{one}
\end{eqnarray}
Here $x,y,z$ are the components of the interatomic separation 
$\mathbf{r}$ and 
$\hbar\Gamma_{ray} =  \alpha^{2}  q^{3} I / 6 \pi  c \varepsilon_{0}^{2}$ 
is the single-atom (spontaneous) Rayleigh scattering rate,
which is proportional to the laser intensity $I$ and to
$\alpha^{2}$, i.e.\ to the saturation
parameter that scales as the inverse square of the detuning
from the nearest atomic resonance \cite{odell2000}.
We assume in the following that the condensate contains
many atoms per cubic wavelength to ensure
the validity of the mean-field approach.
The condensate is taken to be at zero temperature
and the rate of heating due to incoherent 
Rayleigh scattering  is shown to be negligible.

The Gross-Pitaevskii equation \cite{dalfovo99}
for the order parameter $\Psi ({\bf r},t)$ can be obtained starting from
\be
i \hbar  \frac{\partial \Psi }{\partial t} =
\frac{\delta }{\delta \Psi^* } H_{\mathrm{tot}}\;.
\label{ggp}
\ee
Here the total mean-field energy functional 
$H_{\mathrm{tot}} = H_{\mathrm{kin}} + H_{\mathrm{ho}}+
H_{\mathrm{dd}} + H_{\mathrm{scat}}$ consists of:
(a) the kinetic energy,  
$H_{\mathrm{kin}}=\int \,d{\bf r} (\hbar^2/2m)|{\bf \nabla}\Psi|^2$;
(b) the harmonic-trap energy, that can be due to the focus size of the beam,
 $H_{\mathrm{ho}}=\int d{\bf r} \,
V_{\mathrm{ho}}\,n(\br) $, where 
$n = |\Psi({\bf r},t)|^{2}$ is the atomic density
and $V_{\mathrm{ho}} = m\omega_{r}^2 \rho^2 / 2  + m\omega_{z}^2 z^2 / 2$ 
is the harmonic cylindrically symmetric potential,
where $\rho^2 = x^2 + y^2$;
(c) the mean-field energy due to the short-range ($r^{-6}$) 
van der Waals interaction, which will be treated 
within the delta-function pseudo-potential approximation,
$H_{\mathrm{scat}} = (2 \pi a \hbar^{2}/m) \int d{\bf r} \, n(\br)^2 $,
where $a$ is the s-wave scattering length,
and (d) 
the electromagnetically-induced mean-field energy,
$ H_{\mathrm{dd}} =   (1/2) \int d\br\, d\br' \, V_{\mathrm{dd}}(\br-\br')
\, n(\br) n(\br') $ 
which corresponds to the electrostriction energy of the medium
\cite{pita} to lowest order in the $\alpha n / \varepsilon_{0}$.
The mean radii of the condensate 
can be estimated using the Gaussian ansatz
$\Psi(\rho,z)=N^{1/2 } \exp \left(  - \rho^2/ 2w_r^2 - z^2/2 w_z^2 \right)
/ \pi^{3/4} w_z^{1/2} w_r $.
The variational parameters  $w_z$ and $w_r$, which are related to the mean
quadratic radii through $(\Delta x)^2  
\equiv \langle x^2 \rangle = w_r^{2}/2 $ and 
$(\Delta z)^2  \equiv \langle z^2 \rangle  = w_z^{2}/2 $,
are obtained by minimizing $H_{\mathrm{tot}}$.

A self-bound condensate is then found to exist in the 
Thomas-Fermi (TF) limit (negligible kinetic energy).
External confinement is not necessary and the values of $w_{r}$ and 
$w_{z}$ are finite due only to 
the presence of a \emph{plane-wave} laser.
The laser intensities are below the threshold of the instability 
caused by the $1/r^3$ part of the dipole-dipole potential
(\ref{one}) (static field limit) \cite{lewensteinr^3},
$ I \le 12 \pi \hbar^2  c \epsilon_0^2 a / m \alpha^2 $,
and can be, in principle, arbitrarily small.
Here and henceforth we normalize the intensity to 
$8 \pi \hbar^2  c \epsilon_0^2 a / m \alpha^2$
and the spatial scales are normalized to the laser wavelength 
$\lambda_{L}=2\pi/q$.

Figure 1 shows the variational values of the mean condensate radii  
in the TF-limit for a plane wave laser and in the 
absence of external confinement. 
For $I \ll 1$ the radii of the condensate are given by
$\Delta x  = 0.1125 \, I^{-1/2}$ and $\Delta z  = 0.7847 \,I^{-1}$.
We note that for laser intensity $I > 0.1$
the  condensate is strongly confined in the radial direction with 
$\Delta x \sim 0.2 $, and less confined in the longitudinal
direction, with typical size larger or of the order of the wavelength,
$\Delta z >  0.6$.

\begin{figure}[htbp]
\begin{center}
\centerline{\epsfig{figure=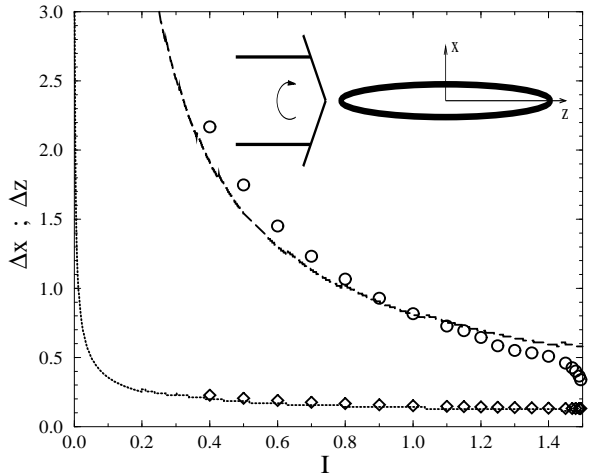,width=9cm,height=7.3cm}}
\end{center}
\vspace{-5ex}
\caption{
Inset---the laser beam and condensate geometry.
Mean radii (normalized to $\lambda_{L}$)
of the self-bound condensate versus the intensity $I$ 
(normalized  to $8 \pi \hbar^2  c \epsilon_0^2 a / m \alpha^2$) 
in the TF-limit.
Long-dashed (dotted) lines represent the expectation value of 
$\Delta z$ ($\Delta x$)
obtained from the Gaussian ansatz.
Circles (diamonds) represent the expectation value of 
$\Delta z$ ($\Delta x$)
calculated from the quasi-1D ansatz (3).}
\label{fig:1}
\end{figure}

The large extension of the condensate along the longitudinal ($\hat{z}$) 
axis means that the oscillatory long-range behavior of
the potential (\ref{eq:ztpot}) can become manifest in this direction.
We study the effect of these oscillations using the ansatz
\be
\Psi(\rho,z) =  \psi(z) \,
\exp \left(  - \rho^2 / 2 w_r^2 \right) / \pi^{1/2} w_r 
\label{eq:ansatz2}
\ee
in order to reduce the problem to a quasi-one dimensional (1D) one.
We then obtain $\psi(z)$, as well as $w_r$, by minimizing numerically
the mean-field energy $H_{\mathrm{tot}}$.
The numerical solutions for the ground state, based on (\ref{eq:ansatz2}), 
yield radial and longitudinal sizes of the \emph{self-confined} condensate 
($H_{\mathrm{ho}}=0)$ that are in good agreement with those based on the 
Gaussian ansatz, as displayed in Fig 1.

The electromagnetically-induced mean-field energy,
expressed through the linear density $n_{\mathrm{1D}} = |\psi(z)|^{2}$,
takes the 1D-reduced form $ H_{\mathrm{dd}}^{\mathrm{1D}} = $ $  (1/2) 
\int dz\, dz'\, V^{\mathrm{1D}}_{\mathrm{dd}}(z-z') 
$ $\, n_{\mathrm{1D}}(z) n_{\mathrm{1D}}(z') $,
where the induced dipole-dipole  potential, reduced to 1D, is given by
\be
V^{\mathrm{1D}}_{\mathrm{dd}}(z) = \frac 1{2\pi w_r^2}\int
d^2\rho \exp \left(  - \frac{\rho^2}{ 2 w_r^2} \right) 
V_{\mathrm{dd}}(\rho,z) \; .
\ee
The 1D-reduced interaction contains an attractive singular part
$ \left. V^{\mathrm{1D}}_{\mathrm{dd}}(z) \right| _{sing.} = 
- (\hbar \Gamma_{ray} / q^{3} w_r^2) \,
 \delta(z)  $. 
For $I<1.5$ the singular part of the reduced interaction 
is balanced by a similar repulsive term 
$V^{\mathrm{1D}}_{\mathrm{scat}}(z) = (2 a \hbar^{2} / m w_r^2) \delta(z)$ 
proportional to the (positive) s-wave scattering length.
Figure 2 shows the non singular part of $V^{\mathrm{1D}}_{\mathrm{dd}}$ 
for three different values of the variational parameter $w_r$.
For large values of $z$ the 1D-reduced potential 
oscillates as $V^{\mathrm{1D}}_{\mathrm{dd}}(z) \approx 
- (3\hbar \Gamma_{ray}/2q) \cos(q z)^2/z$,
giving rise to characteristic
attractive logarithmic singularities in the Fourier transform 
at $k_z=0$ and $k_z=\pm 2 q$ (Fig. 2b).

\begin{figure}[htbp]
\begin{center}
\centerline{\epsfig{figure=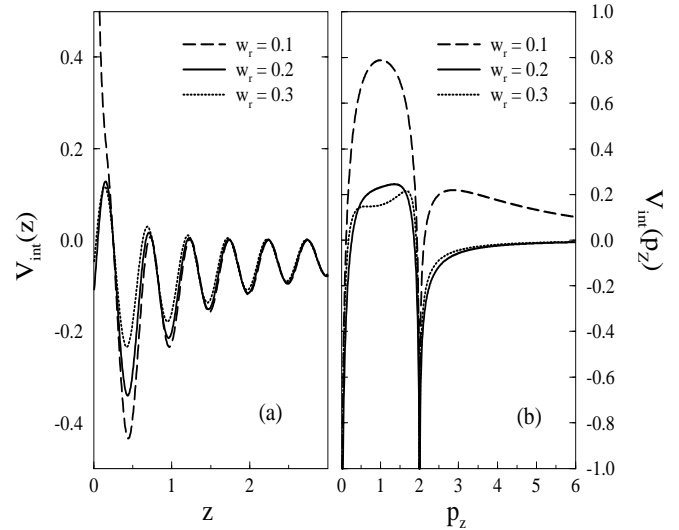,width=8.5cm,height=8.0cm}}
\end{center}
\vspace{-5ex}
\caption{
(a)  the induced 1D dipole-dipole interaction
in units of $\hbar \Gamma_{ray}$,
as a function of $z$ (normalized to $\lambda_L$),
(b)  its Fourier transform.
The dashed, solid and dotted lines correspond to increasing values of the 
variational parameter $w_{r}=0.1$, $0.2$ and $0.3$, respectively.
}
\label{fig:veff}
\end{figure}

In the TF-limit, the Bogoliubov dispersion relation
for the condensate \cite{dalfovo99}
is proportional to the square root of the Fourier transform of the
interparticle potential. In a sense then, the 
$ k_z = \pm 2 q $ singularities give rise to a roton-like minimum (cf.
He-II) in the dispersion relation. 
When the roton minimum
touches the zero frequency axis there is no energy cost to the
system adopting a density modulation in its ground state,
and a 'supersolid' is created \cite{pomeau94} (see below).

A new feature appearing in the TF-limit 
is the formation of \emph{single condensate droplets}
of size less than $\approx .25 \lambda_{L}$ located at an approximately
regular distance of $\lambda_L/2$.
The variational values for ground state energies based on (\ref{eq:ansatz2})
are substantially below the ones obtained by the Gaussian ansatz,
as displayed in Fig. 3a, and indicate that such density modulation 
is quite stable as its energy reduction is of order of the 
Gaussian variational energy.
The effect of density modulation arises as
the field back-scattered  by the condensate interferes with
the incident field, thus modulating the total laser intensity and 
consequently creating a series of attractive dipole traps along the 
longitudinal axis that are separated by half a wavelength
(see  thick curve in Fig. 4).
In this limit, the single condensate droplets become more and more confined 
and their number is reduced,
when approaching the threshold of the $1/r^3$ instability ($I=1.5$).

\begin{figure}[htbp]
\begin{center}
\centerline{\epsfig{figure=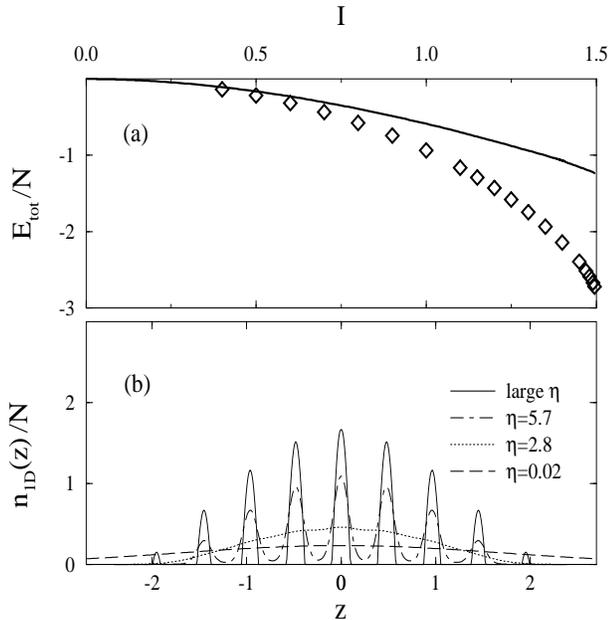,width=9cm,height=8.5cm}}
\end{center}
\vspace{-5ex}
\caption{
(a) TF mean-field energies per particle (normalized  to $\eta E_R$)
versus the intensity $I$ 
(normalized  to $8 \pi \hbar^2  c \epsilon_0^2 a / m \alpha^2 $)
as obtained from the Gaussian ansatz (solid-line)
and from the ansatz (3) (diamonds).
(b) Kinetic energy effects: longitudinal equilibrium densities 
for different values of $\eta$ and for $I=1$.
A radial external confinement is used to keep
the radial size of the condensate constant and equal to that obtained for 
$\eta$ very large without confinement.
}
\label{fig:3}
\end{figure}

The changes in the density profile as the kinetic energy becomes non-negligible
can be discussed in terms of the dimensionless parameter 
\be
\eta = N a / \lambda_{\mathrm L} \;,
\ee
which is approximately the ratio between 
the TF-ground state energy per particle for $I = 1$ 
and the recoil energy $E_R = \hbar^2 q^2 / 2 m $ (see Fig. 3a).
When the product $\eta \, I $ decreases and becomes comparable to one, 
the radii of the condensate both increase with respect to those obtained
in the TF-limit. 
This can be understood comparing the TF-energy, 
$E_{tot}  \sim  N \, \eta \, E_R \, I^2$,
with the kinetic energy, $E_{kin} \sim N \,E_R \, I$, 
estimated using the TF-Gaussian solutions.

We can compensate the kinetic energy pressure in the radial direction,
when necessary,
by adding external radial confinement, e.g. by an appropriate
choice of the focus of the laser beam.
Fig. 3b  shows the results of fixing the radial width
$w_r$ and decreasing the parameter $\eta$.
As soon as the kinetic energy associated with the density modulation,
that is larger as $N \,E_R $,
exceeds the energy reduction caused by density modulation, 
e. g.  $\eta I^2 \sim 1$,
we find the oscillatory pattern to be washed out.
For $\eta$ smaller than $2.8$ at $I=1$, 
the condensate is still self-confined along the longitudinal direction,
with a similar value of the longitudinal size $\Delta z$,
but the density oscillations are almost washed out.
Only for much smaller values of $\eta$, for instance $\sim 0.02$ at $I=1$,
does the kinetic energy also affect  $\Delta z$
(long dashed line in Fig. 3b).

Remarkably, when $\eta I^2 \sim 1$, 
the single condensate droplets overlap, self-organizing into a 
\emph{spatially-coherent long-range density modulation} ---
a Bose ``supersolid'' \cite{pomeau94} is formed.
In this novel regime the atoms are coherently distributed among the wells
allowing the establishment of a phase between the overlapping condensates.
An example is given by the dot-dashed curve in Fig. 3b.

Using the Gaussian ansatz in its time-dependent form \cite{perezg96}
we then obtain analytically in the small intensity limit
the values for the radial and longitudinal oscillation frequencies
for a self-bound condensate in absence of the density modulation 
($I \ll \eta I^2 \ll 1$):
$\Omega_r^2 = 2.88  \, (E_R/\hbar)^2 \, (I^3 \eta)$
and
$\Omega_z^2 = 0.0148  \, (E_R/\hbar)^2 \, (I^4 \eta)$.
It is noteworthy that, although the condensate radii are much larger than 
the laser wavelength (\emph{far-zone}), the frequency $\Omega_r^2$ of the 
compressional mode along the radial direction is close to
the ``plasma'' frequency $\omega_p^2 = 4 \pi u n(0) / m = 
9.86 \, (E_R/\hbar)^2 \, (I^3 \eta)$
introduced for the condensate in the near-zone 
``self-gravity'' regimes \cite{giova2001b},
$n(0)$ being the peak density 
and $u=(11/10)\,\hbar \Gamma_{ray}/q$  the ``gravitational''  coupling 
of the \emph{near zone} laser-induced interatomic potential $-u/r$ 
\cite{odell2000}.
The longitudinal frequency $\Omega_z$ is expected to be substantially
modified in the presence of the oscillatory pattern.

The spontaneous Rayleigh scattering at a rate $\Gamma_{\mathrm ray}$ from the 
interaction-inducing laser beam leads to heating and depletes the condensate. 
Treating the scattering of photons as an incoherent process,
we expect the total energy of the condensate to increase as  \cite{frule}
\be
\frac{d}{dt}\left(E_{tot}\right)
= 2 \,N \,E_{R}\,  \Gamma_{\mathrm ray} \;.
\label{energyflux}
\ee
The characteristic time in which the condensate may heat up is 
$\tau_{\mathrm heat} = \left((d\,E_{tot}/dt)/|E_{tot}| \right)^{-1}$.
We estimate $\tau_{\mathrm heat} =  0.02145 \,N \,(\hbar/E_R)\,I$,
using the relations for the small intensity limit
obtained using the Gaussian ansatz
($E_{tot} / N = - 0.7189 \, \eta \, E_R \, I^2$),
and compare it with the above expression for $\omega_p$.
The product $\tau_{\mathrm heat} \omega_{p}$, that can be expressed
in terms of the number of atoms per cubic wavelength
as $\tau_{\mathrm heat} \omega_{p}
= 0.11 \,(a/\lambda_L)^{1/2} (\lambda_L^3 n(0))^{3/2} I^{-1/2} $,
proves that heating due to spontaneous Rayleigh scattering events 
does not provide a fundamental limitation on the observability of the 
discussed effects for sufficient density.

For the density modulations discussed above,  most of the 
photons are \emph{nearly elastically} and \emph{coherently} scattered.
Their total cross section is given by 
$N^2 f_{cm}$ times the single-atom cross section,
where the fraction  $f_{cm}$ is defined as an appropriate average over all the 
possible directions of scattering (in the Born approximation)
\be
f_{cm} = \frac 3{2 N^2} \int_{-1}^{1} d(\cos(\theta)) \,(1 + \cos(\theta)^2)
\,\left|n(\bk)\right|^2  
\;.
\ee
Here $n(\bk)$ is the Fourier component of atomic density
corresponding to  $\bk=2q(\sin(\theta)\hat{x}, 0 , \cos(\theta)\hat{z})$, 
the momentum transfered by a single photon in the x-z plane  and $\theta$
is the angle between the incident beam and the scattered direction.
Only the (complementary) fraction 
$(1-f_{cm})$ of the rate of energy change (\ref{energyflux}),
that is related to the incoherent part of the scattering cross-section,  
is absorbed by the interatomic degree of freedom in the center-of-mass
frame, thereby providing a partial suppression of heating.

The center of mass is therefore subject to a constant \emph{radiation force}
that is enhanced by a factor $N f_{cm}$ by the 
density modulation.
Correspondingly the electromagnetic field is strongly coherently
back-scattered (diffracted).
The radiation force can either shift the equilibrium position of the 
condensate, if it is located in a longitudinal trap,
or else accelerate it uniformly. 
In the latter case the scattered light will be Doppler shifted.
The effects described here are due to the \emph{same} matter field 
interactions 
as those responsible for stimulated Rayleigh scattering \cite{born+wolf} and
collective atomic recoil (CARL) \cite{moore1999}.
In fact, $f_{cm}$ will lead to similar effects. 
However, the change of atomic energy, unaccounted for by CARL, is the essence 
of our effects.

A central prediction of this paper is that induced dipole-dipole
forces result in a \emph{stationary} density modulation for the 
condensate ground-state.
Density modulations in a condensate can also occur in the presence of  phonon
excitations, as observed in the experiment by Inouye \textit{et al} 
\cite{inouye} that has  demonstrated the superradiant Rayleigh scattering 
\cite{moore1999} of  a \emph{pulse} of light by an atomic BEC.
However, traveling phonons represent excitations of the system and so are
quite distinct from the ground-state density modulation proposed here.  
The two situations can, for instance, be distinguished by diffracting a 
nearly perpendicular probe laser with z-component of the wavevector,
$k_z \approx 2q$.
Sound waves at frequency $\Omega=v_s 2q$, where $v_s$ is the sound velocity, 
lead to a density modulation of the form $\cos(2qz - \Omega t)$
 so that the first diffraction orders (i.e.\ Brillouin peaks) of 
the probe beam (wavenumber $k$), at angles $\theta \approx \pm 2q/k$,
will be frequency shifted to $ck \pm \Omega$ \cite{born+wolf}. 
By contrast, diffraction from the stationary density modulation 
discussed above will be \emph{elastic},
with no frequency shift.

An example of the experimental conditions required for the predicted effects 
involves the following parameters for a cloud of $N \sim 10^3$ sodium atoms:
a circularly polarized laser beam,
red-detuned by 1.7 GHz from the 3S$_{1/2}$ (F=1) $\rightarrow$
3P$_{3/2}$ (F = 0,1,2) transition, gives the threshold
for $1/r^3$ instability 
at $ \approx 525 \; {\mathrm{mW/cm}}^{2}$ \cite{odell2000}.
For such intensity, the acceleration of the center of mass is approximately
$ \approx  500  N  f_{cm}$ [m/s$^2$], where
$f_{cm}$ is a non-negligible fraction 
(for instance $f_{cm}\sim 0.1$ for $I\sim 1$ and  $f_{cm}\sim 1$ 
for $I\ll 1$).
It is possible to balance this force by combining  longitudinal harmonic
confinement with a magnetic field gradient,
so that the condensate will not accelerate.

To conclude, we have demonstrated a new regime of self-confined and 
self-organized ground-state density modulations in a BEC 
illuminated by a single circular polarized laser beam.
We have shown that non-linear scattering of light 
may arise, even in the small-saturation limit, 
and have estimated the enhancement of the total
``elastic'' cross section.
This regime is inherently possible even for a plane-wave laser,
although it is facilitated by the radial focusing of the beam.

This work was supported by the German-Israeli Foundation (GIF).


\begin{thebibliography}{10}



\bibitem{pita}
{L.D. Landau and E.M. Lifshitz,
\textit{Electrodynamics of continuous media} (Pergamon Press,
New York, 1960)}







\bibitem{giova2001a}
{S. Giovanazzi, D. O'Dell, and G. Kurizki, Phys. Rev. A 
{\bf 63}, 031603(R) (2001).}


\bibitem{odell2000}
{D. O'Dell, S. Giovanazzi, G. Kurizki and V.M. Akulin,
Phys. Rev. Lett. {\bf 84}, 5687 (2000).}


\bibitem{giova2001b}
{S. Giovanazzi, G. Kurizki, I. E. Mazets, and S. Stringari, 
cond-mat/0101310, Europhys. Lett. in press..}





\bibitem{pomeau94}
{Y. Pomeau and S. Rica, {P}hys. {R}ev. {L}ett.,
{\bf 72}, 2426 (1994).}



\bibitem{inouye}
{S. Inouye, A.P. Chikkatur, D.M. Stamper-Kurn, J. Stenger, D.E. Pritchard,
and W. Ketterle, Science {\bf 285}, 571 (1999).}


\bibitem{moore1999}
{M.G. Moore and P. Meystre, Phys. Rev. Lett., {\bf 83}, 5202 (1999);
N. Piovella, R. Bonifacio, B.W.J. McNeil, and G.R.M. Robb, 
Opt. Commun. {\bf 187}, 165 (2001).}


\bibitem{thirunamachandran80}
{T. Thirunamachandran, {M}ol. Phys., {\bf 40}, 393 (1980);
D.P.~Craig and T.~Thirunamachandran,
\textit{Molecular Quantum Electrodynamics} (Academic Press,
London, 1984)}



\bibitem{lewensteinr^3}
{The $r^{-3}$ instability has been discussed for a BEC
with \emph{magnetic} (static) dipole-dipole interactions by:
K. G{\'{o}}ral, K. Rz{\c{a}}{\.{z}}ewski and T. Pfau, 
Phys. Rev. A {\bf 61}, 051601(R) (2000). The
(static) $r^{-3}$ part of the  \emph{optically} induced 
dipole-dipole interaction in an \emph{externally trapped} and thus
stabilized BEC has been studied by:
L. Santos, G.V. Shlyapnikov, P. Zoller and M. Lewenstein,
Phys. Rev. Lett. {\bf 85}, 1791 (2000).}


\bibitem{dalfovo99}
{L.{P}. {P}itaevskii, {S}ov. {P}hys. {JETP}, {\bf 13}, 451 (1961);
 {E}.{P}. {G}ross, Nuovo Cimento {\bf 20}, 454 (1961);
 J. Math. Phys. {\bf 4}, 195 (1963);
F. Dalfovo, S. Giorgini, L.~{P}.~{P}itaevskii and S. Stringari,
Rev. Mod. Phys. {\bf 71},  463	(1999); 
A.S. Parkins and D.F. Walls, {P}hys. {R}ep. {\bf 303}, 2 (1998).}




\bibitem{perezg96} 
V. M. PerezGarcia, H. Michinel, J. I. Cirac, et al.,
Phys. Rev. Lett. {\bf 77}, 5320 (1996). 



\bibitem{frule}
{The analogous expression for fast-particle scattering  
can be found in: D. Pines, and P. Nozieres,
\textit{The Theory of  Quantum Liquids} (Benjamin, 1966)}




\bibitem{born+wolf}
{M. Born and E. Wolf, \textit{Principles of Optics} (Pergamon, 1980);
R. Boyd, \textit{Nonlinear Optics} (Academic Press, 1992)}



\end{thebibliography}
\end{document}